\documentclass[onecolumn,amsmath,amssymb,pra,groupedaddress,nobibnotes,showpacs]{revtex4}
\usepackage{graphicx}

\begin{document}
\title{Scalable scheme for entangling multiple ququarts using linear optical elements}

\author{So-Young Baek }
\email{simply@postech.edu}
\affiliation{Department of Physics, Pohang University of Science and Technology (POSTECH), Pohang, 790-784, Korea}

\author{Yoon-Ho Kim}
\email{yoonho@postech.edu}
\affiliation{Department of Physics, Pohang University of Science and Technology (POSTECH), Pohang, 790-784, Korea}

\date{to appear in Phy. Lett. A (2007), doi:10.1016/j.physleta.2007.05.033}

\begin{abstract}
We report a scalable linear optical scheme for generating entangled states of multiple ququarts in which the individual single-ququart state is prepared with the biphoton polarization state of frequency-nondegenerate spontaneous parametric down-conversion. The output state is calculated with the full consideration of the higher order effect (double-pair events) of spontaneous parametric down-conversion. Scalability to multiple-ququart entanglement is demonstrated with examples:  linear optical entanglement of  three and four individual biphoton ququarts. 
\end{abstract}


\maketitle

%%%%%%%%%%%%%%%%%%%%%%%%%%%%%
\section{Introduction}
%%%%%%%%%%%%%%%%%%%%%%%%%%%%%%%%%

In quantum information, the quantum bit (qubit) is often considered as the basic unit of information processing. Physically, the qubit is equivalent a two-dimensional quantum state, such as, electron spin, photon polarization, two discrete energy levels, etc. Due to the intrinsic quantum nature of the qubit, it has two distinct properties in comparison to the classical bit: quantum superposition (for single qubit) and quantum entanglement (for multiple qubits). In quantum information, these distinct quantum properties of the qubit are harnessed to perform certain information processing and communication tasks better than using classical means.

The dimension of the quantum system, however, need not be fixed at two. Continuous variable (i.e., infinite dimensional) quantum states have long been studied in quantum optics to understand squeezing and certain quantum information protocols, such as, quantum teleportation and quantum cryptography have been implemented with continuous variables \cite{ou,furusawa,ralph}. 
It is then natural to consider a high-dimensional discrete quantum state as the potential resource or the basic information unit in quantum information. In fact, recently, there has been much interest in the $D$-dimensional ($D>2$) quantum state (quDit) for potential quantum information applications and for fundamental tests of the quantum theory \cite{kas,Popescu,Chen,peres,lee}.

In the quantum optical approach to quantum information, the qubit is often implemented with a certain easy-to-control two-dimensional degree of freedom of a photon, such as, polarization, path, time-of-arrival, and phase \cite{shih,kim,gisin}. Similarly, the quDit may be implemented with a high-dimensional degree of freedom of a photon, such as, multiple paths, multiple time-bin, angular momentum states, etc. Of particular importance and interest, among the photonic quDit implementation schemes, are reported in Ref.~\cite{bog,chek} and Ref.~\cite{kulik3,kulik4}, in which the qutrit and the ququart states, respectively, are implemented with a pair of photons, rather than with a single photon.

In the recent years, there have been a number of experimental demonstrations of two-quDit entangled states using the high-dimensionality of photonic angular momentum, transverse momentum-position, time-of-arrival, etc. \cite{zeilinger1,zeilinger2,howell,padua,gisin2}. What is common in these seemingly different schemes is that they all utilize the inherent position-momentum or angular-momentum correlations exist between the two photons of spontaneous parametric down-conversion. As a consequence, two entangled quDits are always generated as a pair and the entanglement schemes are, in principle, not scalable to multiple quDits. 

In experimental quantum information research, it is desirable to be able to prepare individual quDit states separately and to have a scheme to entangle such individual quDits in a scalable manner, when needed. This bottom-up approach to entanglement generation is more or less impossible for the quDits reported in Ref.~\cite{zeilinger1,zeilinger2,howell,padua,gisin2} due to the physical nature of the photonic degrees of freedom used to implement the quDit.

Recently, we proposed a linear optical scheme for generating entangled states of two photonic ququarts in which the individual ququart state is made of a pair of photons of different frequencies \cite{baek}. In this paper, we investigate realistic and experimentally realizable two-ququart and multi-ququart entangling schemes in which the single ququart state is based on frequency-nondegenerate two-photon polarization state of spontaneous parametric down-conversion (SPDC) \cite{kulik3,kulik4}. The SPDC process generates a coherent superposition of the vacuum, the two-photon amplitude, the four-photon amplitude, etc \cite{klyshko}. Here, we discuss experimentally relevant (postselected or non-postselected) output states of the linear optical entangling scheme by considering the real SPDC photon states (i.e., coherent superposition of the vacuum, the two-photon, the four-photon, etc. terms) as the input \cite{kok}. We also discuss, with examples, how the beam splitter-based two-ququart entanglement scheme can be scaled up to generate entangled states of multiple biphoton ququarts.

%%%%%%%%%%%%%%%%%%%%%%%%%%%%%%%%%%%%%%%%
\section{Two-photon Ququart}
%%%%%%%%%%%%%%%%%%%%%%%%%%%%%%%%%%%%%%%%

Unlike other schemes that implement quDit states on the multi-dimensional degree of freedom of a single-photon state \cite{zeilinger1,zeilinger2,howell,padua,gisin2},
the ququart, a four-level quantum state, discussed in this paper is based on the two-photon polarization state of frequency-nondegenerate SPDC \cite{kim2}. 

Encoding multi-dimensional quantum states on a photon pair, rather than on a single photon, was first reported in Ref.~\cite{bog,chek,kulik3,kulik4} and there are a number of advantages to this approach. First, since the two-fold coincidence detection is the basic measurement for the pair photon, the scheme is less susceptible to environmental optical noises and the detector dark counts. Second, the photon pair state can be prepared reliably and efficiently with the SPDC process, while there is no efficient and reliable source of single-photon states at the moment \cite{sim}. Third, this two-photon approach is particularly efficient and useful  for the ququart state implementation. The degree of polarization of frequency-nondegenerate two-photon polarization state of SPDC is not invariant under SU(2) transformations \cite{su21,su22}. Therefore, all four biphoton ququart basis states can be easily accessed by a set of waveplates, i.e., linear optically \cite{kulik3,kulik4}. Finally, as we discussed earlier in Ref.~\cite{baek}, it is possible to generate entangled states of two individual ququarts using beam splitters, such as, an ordinary 50/50 beam splitter (BS), a polarization beam splitter (PBS), and a dichroic beam splitter (DBS). In addition, as we shall show in section \ref{multi}, it is possible to generate entangled states of multiple ququarts linear optically.

Let us start by defining the basis states that form the biphoton ququart \cite{kulik3,kulik4}. For the frequency-nondegenerate two-photon polarization state of SPDC, it is not difficult to realize that the two-photon polarization state must belong to the four-dimensional Hilbert space and that the four orthonormal basis states can be written as,
%%%%%%%%%%%%%%
\begin{equation}\label{eq1}
\begin{split}
|H_{\lambda_{1}},H_{\lambda_{2}}\rangle &\equiv |0\rangle, \\
|H_{\lambda_{1}},V_{\lambda_{2}}\rangle  &\equiv |1\rangle, \\
|V_{\lambda_{1}},H_{\lambda_{2}}\rangle & \equiv |2\rangle, \\
|V_{\lambda_{1}},V_{\lambda_{2}}\rangle  &\equiv |3\rangle, 
\end{split}
\end{equation}
%%%%%%%%%%%%%%%%
where subscripts $\lambda_1$ and $\lambda_2$ label two different wavelengths of the photons. $H$ and $V$ refer to the horizontal and vertical polarization states, respectively. It is important to note that these two photons must occupy the same spatial mode.

The most general form of the biphoton single-ququart state is then given by the coherent superposition of the two-photon amplitudes (the ququart basis states) defined in eq.~(\ref{eq1}),
%%%%%%%%%%%%%%%%%%%%%%%
\begin{equation} \label{eq2}
|\psi\rangle = c_{0}|0 \rangle + c_{1}|1 \rangle   + c_{2}|2 \rangle + c_{3}|3 \rangle,
\end{equation}
%%%%%%%%%%%%%%%%%%%%%%%%%
where $c_{j}=|c_{j}|e^{i\varphi_j}$ are the complex probability amplitudes and satisfy the normalization condition $\sum_{j=0}^{3}|c_{j}|^{2}=1$.

%%%%%%%%%%%%%%%%%%
\begin{figure}[t]
\centering
\includegraphics[width=3in]{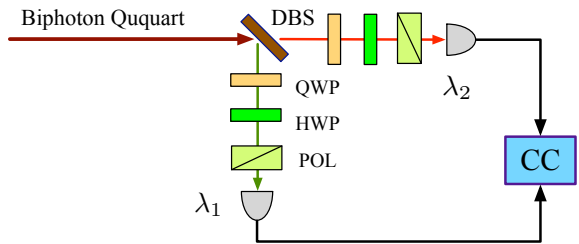}
\caption{The single biphoton ququart detection scheme \cite{kulik3,kulik4}.  The dichroic beam splitter (DBS) splits the two incoming frequency-nondegenerate photons ($\lambda_1$ and $\lambda_2$) into two spatial modes. Polarization projection measurement is performed using a quarter-wave plate (QWP), a half-wave plate (HWP), and a polarizer (POL) on each of the photon. CC is the coincidence counter.} 
\label{fig2}
\end{figure}
%%%%%%%%%%%%%%%%%%

The state in eq.~(\ref{eq2}) can be experimentally implemented using SPDC. For example, two type-I SPDC sources can each be used to prepare the ququart basis states $|H_{\lambda_{1}},H_{\lambda_{2}}\rangle$ and $|V_{\lambda_{1}},V_{\lambda_{2}}\rangle$  and two type-II SPDC sources for each of the basis states $|H_{\lambda_{1}},V_{\lambda_{2}}\rangle$ and $|V_{\lambda_{1}},H_{\lambda_{2}}\rangle$. Coherent combinations of these four SPDC amplitudes will then make up the general single-ququart state in eq.~(\ref{eq2}) and in this way we have the full control of the coefficients in eq.~(\ref{eq2}). Experimentally, however, this is a bit inconvenient and complicated as coherently combining (i.e., without any distinguishing temporal and spectral information) type-I and type-II SPDC is certainly not an easy problem due to the group velocity mismatches and the two-photon bandwidth differences between the two phase matching conditions of SPDC. A simpler procedure would be to generate one basis state in eq.~(\ref{eq1}) using either type-I or type-II SPDC and then to transform the state to a more complex single-ququart superposition state by using a set of waveplates, as reported in Ref.~\cite{kulik3,kulik4}.

To detect the biphoton ququart state, the biphoton ququart detection scheme shown in Fig.~\ref{fig2} can be used \cite{kulik3,kulik4}. As we can see from eq.~(\ref{eq1}), the two photons have the different frequencies $\lambda_1$ and $\lambda_2$ so that they can be easily separated into two spatial modes by using a dichroic beam splitter. The polarization measurement is performed using a quarter-wave plate, a half-wave plate, and a polarizer on each of the now separated single-photons and the coincidence events between the two detectors are recorded at the coincidence counter to obtain the joint polarization measurement output. To completely characterize the single biphoton ququart state (or the density matrix), total of sixteen joint polarization measurements should be performed \cite{kulik3,kulik4,james}.

%%%%%%%%%%%%%%%%%%%%%%%%%%%%%%%%%%%%%%%%%%%%%%%
\section{Entangling schemes for two biphoton ququarts}
%%%%%%%%%%%%%%%%%%%%%%%%%%%%%%%%%%%%%%%%%%%%

To discuss the linear optical two-ququart entangling scheme, let us start with two independent biphoton ququarts $|\psi\rangle_a$ and $|\psi\rangle_b$ of the form,
%%%%%%%%%%%%%%%%%%%%%%%%%%%%
\begin{equation}\label{eq3}
\begin{split}
|\psi\rangle_{a} &= c_{0}|0 \rangle_{a} + c_{1}|1
\rangle_{a} + c_{2}|2 \rangle_{a} + c_{3}|3 \rangle_{a},\\
|\psi\rangle_{b} &= c_{0}'|0 \rangle_{b}+ c_{1}'|1
\rangle_{b}+ c_{2}'|2 \rangle_{b}+ c_{3}'|3 \rangle_{b}. 
\end{split}
\end{equation}
%%%%%%%%%%%%%%%%%%%%%%%%%%%%%%%

\begin{figure}[t]
\centering
\includegraphics[width=3.4in]{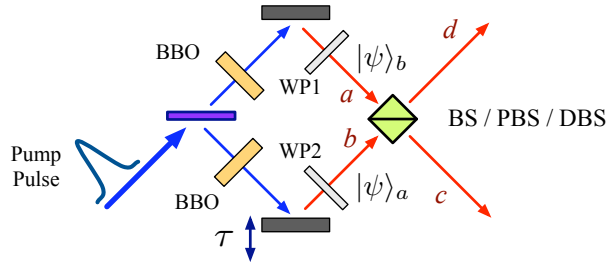}
\caption{The proposed linear optical two-ququart entanglement scheme. Waveplates WP1 and WP2 transform the biphoton states to the desired ququart states $|\psi\rangle_a$  and $|\psi\rangle_b$. Three different types of beam splitters are investigated: an ordinary 50/50 beam splitter (BS), a polarizing beam splitter (PBS), and a dichroic beam splitter (DBS).}
\label{fig1}
\end{figure}

In the linear optical two-ququart entanglement scheme, the entangled state of two biphoton ququarts is generated by interfering them at a beam splitter. The schematic of the proposed linear optical two-ququart entanglement experiment is shown in Fig.~\ref{fig1}. A femtosecond uv laser pulse is split into two and pump the SPDC generating BBO crystals. Using the waveplates WP1 and WP2, the biphoton state is transformed to the desired ququart states $|\psi\rangle_a$  and $|\psi\rangle_b$ and they are made to overlap (or interfere) at the beam splitter. Here we consider three different types of beam splitters, an ordinary 50/50 beam splitter (BS), a polarizing beam splitter (PBS), and a dichroic beam splitter (DBS). Since Shih-Alley and Hong-Ou-Mandel type quantum interference is used to generate entangled states of the two ququarts, it is necessary that the SPDC process is pumped by a ultrafast laser \cite{kim2,hom,sa}. The delay $\tau$ adjusts the overlap of the two ququarts at the beam splitter, hence controlling temporal distinguishability. 

%%%%%%%%%%%%%%%%%%
%%%%%%%%%%%%%%%%%%

The biphoton state $|\psi\rangle_{out}$ at the output modes $c$ and $d$ of the beam splitter is then investigated for three different types of beam splitters, an ordinary 50/50 beam splitter (BS), a polarizing beam splitter (PBS), and a dichroic beam splitter (DBS). Assuming that the beam splitter is loss-less, we can describe it as a unitary matrix acting on incoming photons. Therefore, once the beam splitter unitary matrix is fully determined, the output state $|\psi\rangle_{out}$ can be readily calculated.  

Note that we are only interested in the biphoton states that can be expressed in the ququart basis states in eq.~(\ref{eq1}). This is justified because, to detect a single biphoton ququart, it is necessary to perform the two-photon coincidence measurement shown in Fig.~\ref{fig2}. Any biphoton amplitudes, at the output modes of the beam splitter, that cannot be expressed in the biphoton ququart basis states in eq.~(\ref{eq1}), therefore, do not result in detection events and can be ignored.

The input two-ququart state to the beam splitter, as shown in Fig.~\ref{fig1}, is obviously a product states. So, at first sight, one may simply write the input state to the beam splitter as, 
%%%%%%%%%%%%%%%%%%%%%%%
\begin{equation}\label{input} 
|\psi\rangle_{in} = |\psi\rangle_{a}\otimes |\psi\rangle_{b}.
\end{equation}
%%%%%%%%%%%%%%%%%%%%%

Equation (\ref{input}), which assumes that each SPCD source emits exactly one pair of photons, however, is incorrect in general for the interference experiment involving two separate SPDC sources. The SPDC process, in reality, leads to coherent superposition of the vacuum, two-photon (single pair), four-photon (double-pair), and higher-order terms \cite{klyshko,kok}. The individual ququart states entering the BS, therefore, should be modified as, 
%%%%%%%%%%%%%%%%%%%%
\begin{equation}\label{SPDC_ququart}
\begin{split}
|\psi\rangle_{a}^{SPDC} &= |vac\rangle_{a} + \eta |\psi\rangle_{a} + \eta^{2} |\psi\rangle_{a}\otimes|\psi\rangle_{a} +\cdots, \\
|\psi\rangle_{b}^{SPDC} &= |vac\rangle_{b} + \eta |\psi\rangle_{b} + \eta^{2} |\psi\rangle_{b}\otimes|\psi\rangle_{b} +\cdots,
\end{split}
\end{equation}
%%%%%%%%%%%%%%%%%%%%%
where $|vac\rangle$  is the vacuum,  $\eta$ is related to the two-photon (ququart) event efficiency, and $|\psi\rangle_a$  ($|\psi\rangle_b$) is as defined in eq.~(\ref{eq1}). Note that the first order terms of $\eta$ are the desired two-photon event terms. In two-photon coincidence experiments,  higher order terms ($\eta^2$ and higher) are normally dropped since $\eta \ll 1$. 

Clearly, the input state to the beam splitter in eq.~(\ref{input}) should also be modified as
%%%%%%%%%%%%%%%%
\begin{equation}\label{real_in}
|\psi\rangle_{in}^{SPDC} = |\psi\rangle_{a}^{SPDC} \otimes |\psi\rangle_{b}^{SPDC}.
\end{equation}
%%%%%%%%%%%%%%%%%
Substituting eq.~(\ref{SPDC_ququart}) into eq.~(\ref{real_in}), we find that, 
%%%%%%%%%%%%%%%%
\begin{equation}\label{real_in2}
|\psi\rangle_{in}^{SPDC} \approx \eta^2 (|\psi\rangle_a |\psi\rangle_b +|\psi\rangle_a |\psi\rangle_a +|\psi\rangle_b |\psi\rangle_b ),
\end{equation}
%%%%%%%%%%%%%%%%%
where $|\psi\rangle_a |\psi\rangle_a$, for example, is a short-hand notation for $|\psi\rangle_a \otimes |\psi\rangle_a \otimes |vac\rangle_b$ which refers to the event that four-photons are generated at the SPDC source $a$ while no photons are generated at the SPDC source $b$. Here, we have dropped the non-detectable vacuum component $|vac\rangle$ as well as amplitudes that are higher order than $\eta^2$.

Note now that we can no longer ignore the four-photon event of the SPDC, the second and the third terms in eq.~(\ref{real_in2}), as they have the same probability as the desired input state, the first term of eq.~(\ref{real_in2}).

%%%%%%%%%%%%%%%%%%%%%%%%%%%%%%%%%%%%%%
\subsection{Ordinary 50/50 beam splitter}\label{sec:bs}
%%%%%%%%%%%%%%%%%%%%%%%%%%%%%%%%%%%

For an ordinary 50/50 beam splitter (BS), the unitary transformation is obtained from the welll-known beam splitter input-output relation, 
%%%%%%%%%%%%%%%%%%%%
\begin{equation}\label{BSinout}
\begin{split}
\hat{c}_{jk} &= (\hat{a}_{jk}+ i \hat{b}_{jk})/\sqrt{2}, \\ 
\hat{d}_{jk} &= (i \hat{a}_{jk}+ \hat{b}_{jk})/\sqrt{2},
\end{split}
\end{equation}
%%%%%%%%%%%%%%%%%%%%%%
where $\hat{c}$, for example, is the annihilation operator for a
photon in mode $c$. The subscripts $j$ and $k$ refer to the polarization mode ($|H\rangle$ or $|V\rangle$) and the frequency mode ($\lambda_{1}$ or $\lambda_{2}$) of the photon, respectively.

For a single-photon at the input mode of the BS, as in Fig.~\ref{fig1}, there are four possible states (two spatial modes $a$ or $b$ and two polarization modes $H$ or $V$). We can therefore define the single-photon states as follows \cite{baek}
%%%%%%%%%%%%%%%%%%%%
\begin{equation}\label{singlephoton}
\begin{split}
\hat{a}_{Hk}^{\dagger}|0\rangle = (1,0,0,0)^T, \quad
\hat{a}_{Vk}^{\dagger}|0\rangle = (0,1,0,0)^T,\\
\hat{b}_{Hk}^{\dagger}|0\rangle = (0,0,1,0)^T, \quad
\hat{b}_{Vk}^{\dagger}|0\rangle = (0,0,0,1)^T.
\end{split}
\end{equation}
%%%%%%%%%%%%%%%%%%%%%%%%

Equation (\ref{singlephoton}) then forms the basis set for the unitary transformation $U_{\lambda_1}^{BS}$ of the single-photon of frequency $\lambda_1$  due to the BS. Using eq.~(\ref{BSinout}) and eq.~(\ref{singlephoton}), we find that $U_{\lambda_1}^{BS}$ is given as,
%%%%%%%%%%%%%%%%%%%%%%%%%
\begin{equation}
    U_{\lambda_{1}}^{BS}= \frac{1}{\sqrt{2}}
    \left(
      \begin{array}{cccc}
        1 & 0 & i & 0 \\
        0 & 1 & 0 & i \\
        i & 0 & 1 & 0 \\
        0 & i & 0 & 1 \\
      \end{array}
    \right).
\end{equation}
%%%%%%%%%%%%%%%%%%%%%%%%%%
Note that $U_{\lambda_{1}}^{BS}=U_{\lambda_{2}}^{BS}$ as the BS unitary transformation does not depend on frequency and polarization states.

Remember now that the ququart is in fact made of a pair of photons. The two-ququart interference-entangling scheme shown in Fig.~\ref{fig1}, therefore, can be understood as a four-photon interference scheme with four photons incident at the input ports $a$ and $b$ of the BS, according to eq.~(\ref{real_in2}). The unitary transformation that describes this effect is then calculated to be,
%%%%%%%%%%%%%%%%%%%%%%%%%%
\begin{equation}\label{BS_unitary}
U^{BS} = U_{\lambda_{1}}^{BS} \otimes U_{\lambda_{2}}^{BS} \otimes
U_{\lambda_{1}}^{BS} \otimes U_{\lambda_{2}}^{BS}.
\end{equation}
%%%%%%%%%%%%%%%%%%%%%%%%%%%%
Equation (\ref{BS_unitary}), a $256 \times 256$ matrix, describes all possible unitary transformations by the BS due to the input state given in eq.~(\ref{input}).

Let us first consider the state transformation of the two-photon term $|\psi\rangle_a|\psi\rangle_b$ in eq.~(\ref{real_in2}). Using eq.~(\ref{BS_unitary}), the output state of the BS for the state $|\psi\rangle_a|\psi\rangle_b$ is calculated to be, 
%%%%%%%%%%%%%%%%%%
\begin{equation}\label{BS_out1}
\begin{split}
& |\psi\rangle_a |\psi\rangle_b  \xrightarrow{BS}  \frac{1}{4}
(c_{0}c_{3}'+c_{3}c_{0}'-c_{1}c_{2}'-c_{2}c_{1}')
\\ &~~~~~ \times (|0\rangle_{c} |3 \rangle_{d} + |3\rangle_{c} |0 \rangle_{d} - |1\rangle_{c} |2 \rangle_{d}- |2\rangle_{c} |1 \rangle_{d})\\
&~~~~~ \equiv |\psi\rangle_{2}^{BS}.
\end{split}
\end{equation}
%%%%%%%%%%%%%%%%
In equation (\ref{BS_out1}), there are total of 64 additional four-photon amplitudes that cannot be expressed in the ququart basis states defined in eq.~(\ref{eq1}). Since these amplitudes cannot be detected by the ququart detection scheme shown in Fig.~\ref{fig2}, they have been dropped out. The other 176 amplitudes simply vanishes due to the Hong-Ou-Mandel type destructive quantum interference. (Note that, since the state is pure, the output state is described by a $256 \times 1$ column vector.) 

The output states resulting from the four-photon terms, the second and the third terms in eq.~(\ref{real_in2}), can be evaluated in a similar way. After a lengthy calculation, we find that the output state for the second term in eq.~(\ref{real_in2}) is  given as, 
%%%%%%%%%%%%%%%%%
\begin{widetext}
\begin{equation}\label{double}
\begin{split}
|\psi\rangle_a |\psi\rangle_a |vac\rangle_b \xrightarrow{BS} 
& -  c_0^2 |0\rangle_c |0\rangle_d -  c_0 c_1 |0\rangle_c |1\rangle_d -  c_0 c_2 |0\rangle_c |2\rangle_d - \frac{1}{2} (c_0 c_3 + c_1 c_2) |0\rangle_c |3\rangle_d\\
& -  c_0 c_1 |1\rangle_c |0\rangle_d - c_1^2 |1\rangle_c |1\rangle_d - \frac{1}{2} (c_0c_3 + c_1 c_2) |1\rangle_c |2\rangle_d - c_1 c_3 |1\rangle_c |3\rangle_d\\
& -  c_0 c_2 |2\rangle_c |0\rangle_d - \frac{1}{2} (c_0 c_3 + c_1 c_2) |2\rangle_c |1\rangle_d -  c_2^2 |2\rangle_c |2\rangle_d -  c_2 c_3 |2\rangle_c |3\rangle_d\\
& - \frac{1}{2}(c_0 c_3 + c_1 c_2) |3\rangle_c |0\rangle_d -  c_1 c_3 |3\rangle_c |1\rangle_d -  c_2 c_3 |3\rangle_c |2\rangle_d -  c_3^2 |3\rangle_c |3\rangle_d\\
& \equiv |\psi\rangle_4^{BS}.
\end{split}
\end{equation}
\end{widetext}
%%%%%%%%%%%%%%%%%%
Here, similarly to eq.~(\ref{BS_out1}), only the biphoton terms that can be expressed in the ququart basis state defined in eq.~(\ref{eq1}) are shown. Total of 192 biphoton amplitudes are omitted (postselected out) in eq.~(\ref{double}) as they do not trigger the four-fold coincidence circuit which is at the heart of the two-ququart detection scheme. 

The third term in eq.~(\ref{real_in2}), has the identical output state as in eq.~(\ref{double}) except that all the coefficients are now primed, following the definition of $|\psi\rangle_b$ in eq.~(\ref{eq3}). We shall define this output state as $|\psi\rangle_4^{BS'}$.

The final output state of the BS postselected by the four-fold coincidence measurement, due to the input state in eq.~(\ref{real_in2}), should then be expressed as the superposition of eq.~(\ref{BS_out1}) and eq.~(\ref{double}),
%%%%%%%%%%%%%%%%%
\begin{equation}\label{BS_final}
|\psi\rangle_{out}^{BS} = |\psi\rangle_2^{BS} + |\psi\rangle_4^{BS}+|\psi\rangle_4^{BS'}.
\end{equation}
%%%%%%%%%%%%%%%%%

Clearly, the output state is a rather complicated two-ququart entangled state. Experimentally, one may want to generate a more manageable two-ququart entangled state and, to do so, it is necessary to make appropriate choices of the coefficients in eq.~(\ref{eq3}).

As an example of the two-ququart entangled state, consider the case where $c_0=1$ and $c'_3=1$ in eq.~(\ref{eq3}). Then the real input state to the BS is given from eq.~(\ref{real_in2}) as
%%%%%%%%%%%%%%%%
\begin{equation}
|\psi\rangle_{in}^{SPDC} = |0\rangle_a |3\rangle_b + |0\rangle_a |0\rangle_a + |3\rangle_b |3\rangle_b.
\end{equation}
%%%%%%%%%%%%%%%%%%%%%%
The second and the third terms come from the double-pair events of SPDC. Given the above state as the input to the BS, the postselected output state is then found to be
%%%%%%%%%%%%%%%%
\begin{equation}
\begin{split}
|\psi\rangle_{out} & = - |0\rangle_{c} |0\rangle_{d}-|3\rangle_{c} |3 \rangle_{d} + \frac{1}{4}(|0\rangle_{c} |3 \rangle_{d} \\
& + |3\rangle_{c} |0 \rangle_{d} - |1\rangle_{c} |2\rangle_{d}- |2\rangle_{c} |1 \rangle_{d}),
\end{split}
\end{equation}
%%%%%%%%%%%%%%%%%%
which is a rather complicated-looking two-ququart entangled state.

%%%%%%%%%%%%%%%%%%%%%%%%%%%%%%
\subsection{Polarizing beam splitter}\label{sec:pbs}
%%%%%%%%%%%%%%%%%%%%%%%%%%%%%%%%%

To investigate two-ququart entangling properties of the polarizing beam splitter (PBS), we first need to determine the proper form of the PBS transformation. Following the spatial mode structure in Fig.~\ref{fig1}, we have,
%%%%%%%%%%%%%%%%%%%%%
\begin{equation}
\begin{split}
\hat a_{H k} & \rightarrow  \hat c_{H k}, \quad \quad
\hat a_{V k}  \rightarrow \hat d_{V k}, \\
\hat b_{H k} & \rightarrow \hat d_{H k}, \quad \quad
\hat b_{V k} \rightarrow \hat c_{V k},
\end{split}
\end{equation}
%%%%%%%%%%%%%%%%%%%
where the subscript $k$ refer to the frequency mode  ($\lambda_{1}$ and $\lambda_{2}$) of the photon. 

The PBS unitary transformation for the single-photon of frequency $\lambda_1$ is then given as,
%%%%%%%%%%%%%%%%
\begin{equation}
    U_{\lambda_{1}}^{PBS}=
    \left(
      \begin{array}{cccc}
        1 & 0 & 0 & 0 \\
        0 & 0 & 0 & 1 \\
        0 & 0 & 1 & 0 \\
        0 & 1 & 0 & 0 \\
      \end{array}
    \right).
\end{equation}
%%%%%%%%%%%%%%%%%
Since the PBS unitary transformation does not depend on the frequency mode, $U_{\lambda_{1}}^{PBS}=U_{\lambda_{2}}^{PBS}$. 

The PBS unitary transformation that corresponds to the input state in eq.~(\ref{real_in2}) is therefore calculated to be,
%%%%%%%%%%%%%%%%%%%%%%%%%%%
\begin{equation}\label{pbs_unit}
U^{PBS} = U_{\lambda_{1}}^{PBS} \otimes U_{\lambda_{2}}^{PBS} \otimes U_{\lambda_{1}}^{PBS} \otimes U_{\lambda_{2}}^{PBS},
\end{equation}
%%%%%%%%%%%%%%%%%%%%%%
which is again a $256 \times 256$ matrix.

Applying the PBS unitary transformation in eq.~(\ref{pbs_unit}) to the the two-photon term in eq.~(\ref{real_in2}), we get the following postselected two-ququart state (out of total of 16 amplitudes) in the output modes $c$ and $d$ of the PBS. 
%%%%%%%%%%%%%%%%%%
\begin{equation}\label{PBSout}
\begin{split}
|\psi\rangle_a |\psi\rangle_b  \xrightarrow{PBS} &~ c_{0}c_{0}'|0\rangle_{c} |0 \rangle_{d} + c_{1}c_{1}'|1\rangle_{c} |1 \rangle_{d} \\
& + c_{2}c_{2}'|2\rangle_{c} |2 \rangle_{d} + c_{3}c_{3}'|3\rangle_{c} |3 \rangle_{d}
\end{split}
\end{equation}
%%%%%%%%%%%%%%%%%

For the four-photon amplitudes in eq.~(\ref{real_in2}), the postselected output states (out of total of 16 amplitudes) are given as,
%%%%%%%%%%%%%%%%%%
\begin{equation}\label{PBSout2}
\begin{split}
|\psi\rangle_a |\psi\rangle_a |vac\rangle_b \xrightarrow{PBS} & ~ 2 ( c_0 c_3 + c_1 c_2) |0\rangle_c |3\rangle_d,\\
|vac\rangle_a |\psi\rangle_b |\psi\rangle_b  \xrightarrow{PBS} & ~ 2 ( c'_0 c'_3 + c'_1 c'_2) |3\rangle_c |0\rangle_d.
\end{split}
\end{equation}
%%%%%%%%%%%%%%%%%

Finally, we find that, given the input two-ququart product state in eq.~(\ref{real_in2}), the PBS unitary transformation and the four-photon coincidence postselection gives the  two-ququart entangled state at the output in the form,
%%%%%%%%%%%%%%%%%%
\begin{equation}\label{PBSout4}
\begin{split}
|\psi\rangle_{out}^{PBS} & = c_{0}c_{0}'|0\rangle_{c} |0 \rangle_{d} + c_{1}c_{1}'|1\rangle_{c} |1 \rangle_{d}  \\
& + c_{2}c_{2}'|2\rangle_{c} |2 \rangle_{d}  + c_{3}c_{3}'|3\rangle_{c} |3 \rangle_{d} \\
& + 2 ( c_0 c_3 + c_1 c_2) |0\rangle_c |3\rangle_d \\
& + 2 ( c'_0 c'_3 + c'_1 c'_2) |3\rangle_c |0\rangle_d.
\end{split}
\end{equation}
%%%%%%%%%%%%%%%%%

The output state in eq.~(\ref{PBSout4}), which is not that easy to manipulate experimentally, is the result for the most general superposition input state (pure state) with no coefficients equal to zero. With less general input ququart states, we can obtain a two-ququart entangled state which is experimentally more maneageable. 

Consider the case where $c_1 = c_2=0$ and $c'_1=c'_2=0$ in eq.~(\ref{eq3}). Then the real SPDC input state to the PBS is given as eq.~(\ref{real_in2}) with $|\psi\rangle_{a} = c_{0}|0 \rangle_{a}+c_{3}|3 \rangle_{a}$  and $|\psi\rangle_{b} = c_{0}'|0 \rangle_{b}+c_{3}'|3 \rangle_{b}$. The postselected output state of the PBS then has the simple form
%%%%%%%%%%%%%%%%%%%%%%%%%%
\begin{equation}
\begin{split}
|\psi\rangle_{out} & =  c_{0}c_{0}'|0\rangle_{c} |0 \rangle_{d} +
c_{3}c_{3}'|3\rangle_{c} |3 \rangle_{d} \\
& + 2c_{0}c_{3}|0\rangle_{c}|3\rangle_{d} +
2c_{0}'c_{3}'|3\rangle_{c}|0\rangle_{d}.
\end{split}
\end{equation}
%%%%%%%%%%%%%%%%%%%%%%%%%%%%
Note that, with the superposition of the ququart basis states $|0\rangle$ and $|3\rangle$ as the input, the output two-quauart entangled state also contains only these basis states.

%%%%%%%%%%%%%%%%%%%%%%%%%%
\subsection{Dichroic beam splitter}\label{sec:dbs}
%%%%%%%%%%%%%%%%%%%%%%%%%%%%%

We now consider two biphotons ququarts interfering at a dichroic beam splitter (DBS), see Fig.~\ref{fig1}. The DBS is designed to reflect the $\lambda_1$ photon and to transmit $\lambda_2$ photon, as in the case of the single biphoton ququart detection scheme in Fig.~\ref{fig2}.  

Depending on the frequency ($\lambda_1$ or $\lambda_2$) and the input spatial mode ($a$ or $b$), the single photon entering the DBS is routed as follows,
%%%%%%%%%%%%%%%%
\begin{equation}
\begin{split}
\hat a_{j \lambda_1} &\rightarrow \hat d_{j \lambda_1}, \quad \quad
\hat a_{j \lambda_2} \rightarrow \hat c_{j \lambda_2}, \\
\hat b_{j \lambda_1} &\rightarrow \hat c_{j \lambda_1}, \quad \quad
\hat b_{j \lambda_2} \rightarrow \hat d_{j \lambda_2}.
\end{split}
\end{equation}
%%%%%%%%%%%%%%%%
Here the subscript $j$ refer to the polarization state of a photon ($H$ or $V$).

Since the DBS always reflects $\lambda_1$ photon and transmits $\lambda_2$ photon regardless of the input spatial mode, $\lambda_1$ and $\lambda_2$ photons behave differently at the DBS. For the $\lambda_1$ photon, the unitary transformation matrix is given as,
%%%%%%%%%%%%%%%%%%%%%%
\begin{equation}
U_{\lambda_{1}}^{DBS} =
    \left(
      \begin{array}{cccc}
        0 & 0 & 1 & 0 \\
        0 & 0 & 0 & 1 \\
        1 & 0 & 0 & 0 \\
        0 & 1 & 0 & 0 \\
      \end{array}
    \right),
    \end{equation}
%%%%%%%%%%%%%%
while for the $\lambda_2$ photon, the corresponding unitary matrix is a different one:%%%%%%%%%%%%%%%%%%%%%%
\begin{equation}
U_{\lambda_{2}}^{DBS} =
    \left(
      \begin{array}{cccc}
        1 & 0 & 0 & 0 \\
        0 & 1 & 0 & 0 \\
        0 & 0 & 1 & 0 \\
        0 & 0 & 0 & 1 \\
      \end{array}
    \right),
\end{equation}
%%%%%%%%%%%%%%
These two matrices are simple mathematical representations of the wavelength dependent routing property of the DBS used in this experiment. 

The most general unitary transformation for the DBS that applies to the input four-photon or two-ququart state in eq.~(\ref{real_in2}) is therefore,
%%%%%%%%%%%%%%%%%%%%%%%%%
\begin{equation}\label{DiU}
 U^{DBS} = U_{\lambda_{1}}^{DBS} \otimes U_{\lambda_{2}}^{DBS} \otimes U_{\lambda_{1}}^{DBS} \otimes U_{\lambda_{2}}^{DBS}.
\end{equation}
%%%%%%%%%%%%%%%%%%
Again, this is a $256 \times 256$ matrix.

For the general two-ququart pure state defined in eq.~(\ref{real_in2}) as the input to the DBS, the output state can be determined by using eq.~(\ref{DiU}) and it is given as
%%%%%%%%%%%%%%%%%%
\begin{widetext}
\begin{equation}\label{dbs}
\begin{split}
|\psi\rangle_{out}^{DBS} & =  c_{0}c_{0}'|0\rangle_{c} |0 \rangle_{d}
    + c_{0}c_{1}'|0\rangle_{c} |1 \rangle_{d}
    + c_{0}c_{2}'|2\rangle_{c} |0 \rangle_{d}
    + c_{0}c_{3}'|2\rangle_{c} |1 \rangle_{d}\\
   & +  c_{1}c_{0}'|1\rangle_{c} |0 \rangle_{d}
    + c_{1}c_{1}'|1\rangle_{c} |1 \rangle_{d}
    + c_{1}c_{2}'|3\rangle_{c} |0 \rangle_{d}
    + c_{1}c_{3}'|3\rangle_{c} |1 \rangle_{d} \\
   & +  c_{2}c_{0}'|0\rangle_{c} |2 \rangle_{d}
    + c_{2}c_{1}'|0\rangle_{c} |3 \rangle_{d}
    + c_{2}c_{2}'|2\rangle_{c} |2 \rangle_{d}
    + c_{2}c_{3}'|2\rangle_{c} |3 \rangle_{d}\\
   & +  c_{3}c_{0}'|1\rangle_{c} |2 \rangle_{d}
    + c_{3}c_{1}'|1\rangle_{c} |3 \rangle_{d}
    + c_{3}c_{2}'|3\rangle_{c} |2 \rangle_{d}
    + c_{3}c_{3}'|3\rangle_{c} |3 \rangle_{d}.
\end{split}
\end{equation}
\end{widetext}
%%%%%%%%%%%%%%%%%%%

Equation (\ref{dbs}) has a few interesting properties. First, the 16 terms in eq.~(\ref{dbs}) are all of the non-vanishing amplitudes. Therefore postselection is not necessary if only the two-photon event is present at the input, i.e., only the first term in eq.~(\ref{real_in2}). Clearly, this is not possible with SPDC. However, the non-postselection nature of the DBS two-ququart entanglement scheme could prove to be important if recently reported semiconductor quantum-dot photon pair sources with supressed the multi-pair events become more widely available \cite{akopian}. Second, the second and the third terms in eq.~(\ref{real_in2}) do not result in any detectable two-ququart amplitudes in modes $c$ and $d$ so that they do not contribute to the final postselected amplitudes. With postselection, therefore, the SPDC input state could be considered as the ideal input as in eq.~(\ref{input}) in the case of DBS. Finally, if the input ququart states are formed by equal superposition of all ququart bases states, $ c_{0} = c_{1} = c_{2} = c_{3}$ and $ c_{0}' = c_{1}' = c_{2}' = c_{3}'$, the output state turns out to be a two-ququart product state, i.e., $|\psi\rangle_{out} = |\psi\rangle_{c}\otimes |\psi\rangle_{d}$. 

The DBS can be used to generate a number of interesting two-ququart entangled states starting from rather simple looking two-ququart product states as the input. For example, consider two input ququarts,
%%%%%%%%%%%%
\begin{equation}
\begin{split}
|\psi\rangle_{a} & = \frac{1}{\sqrt{2}} (|0 \rangle_{a}  + e^{i \varphi_1}|3\rangle_{a}), \\
|\psi\rangle_{b} & = \frac{1}{\sqrt{2}} (|0\rangle_{b} + e^{i \varphi_2}|3\rangle_{b}),
\end{split}
\end{equation}
%%%%%%%%%%%%%%
where $\varphi_1$ and $\varphi_2$ are the relative phase terms. 

The DBS output state is a two-ququart entangled state of the form,
%%%%%%%%%%%%%%
\begin{equation}
\begin{split}
|\psi\rangle_{out} & =  \frac{1}{2} (|0\rangle_{c} |0 \rangle_{d}  + e^{i \varphi_1} |1\rangle_{c} |2 \rangle_{d}  + e^{i \varphi_2} |2\rangle_{c} |1 \rangle_{d} \\ & + e^{i (\varphi_1+\varphi_2)} |3\rangle_{c}|3\rangle_{d}).
\end{split}
\end{equation}
%%%%%%%%%%%%%%

%%%%%%%%%%%%%%%%%%%%%%%%%%%%%%%%%
\section{Entangling schemes for multiple ququarts: Examples}\label{multi}
%%%%%%%%%%%%%%%%%%%%%%%%%%%%%%%%

It is not difficult to see that the linear optical scheme for entangling two biphoton ququarts we have discussed so far can, in principle, be scaled up to entangle multiple biphoton ququarts. To discuss this possibility, we consider two simple examples: the three-ququart entanglement and the four-ququart entanglement.

\begin{figure}[tb]
\centering
\includegraphics[width=3.3in]{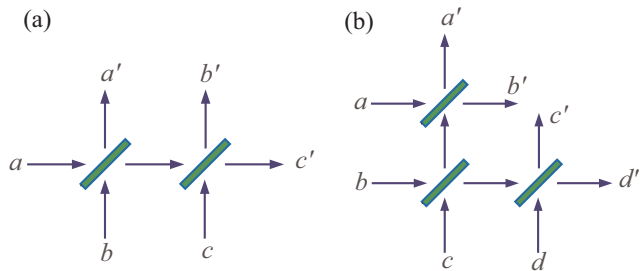}
\caption{Scheme for linear optically entangling three (a) and four (b) biphoton ququarts. To generate the entangled state of $n$ biphoton ququarts, $n-1$ DBS would be needed.} 
\label{fig3}
\end{figure}

First, consider the three-ququart entanglement scheme shown in Fig.~\ref{fig3}(a). The three individual ququarts in the input modes of the DBS are
%%%%%%%%%%%%%%%
\begin{equation}
\begin{split}
|\psi\rangle_a & = |0\rangle_a,\\
|\psi\rangle_b & = |0\rangle_b + |3\rangle_b,\\
|\psi\rangle_c & = |0\rangle_c + |3\rangle_c.
\end{split}
\end{equation}
%%%%%%%%%%%%%%%%%
Two individual ququarts in modes $a$ and $b$ get entangled by the action of the DBS and then one of the subsystem of the two-ququart entangled state becomes entangled with another ququart in mode $c$ with a different DBS. 

As the result, the output state in modes $a'$, $b'$, and $c'$ is given as
%%%%%%%%%%%%%%%%%%%%
\begin{equation}
\begin{split}
|\psi\rangle_{out} & = |0\rangle_{a'} |0\rangle_{b'} |0\rangle_{c'} + |0\rangle_{a'} |2\rangle_{b'} |1\rangle_{c'} \\ &+ |2\rangle_{a'} |1\rangle_{b'} |0\rangle_{c'} + |2\rangle_{a'} |3\rangle_{b'} |1\rangle_{c'},
\end{split}
\end{equation}
%%%%%%%%%%%%%%%%%%%%
which is clearly a three-ququart entangled state. 

The four-ququart entangled state can be generated in a similar way but with one more DBS. Roughly speaking, to entangle $n$ biphoton ququarts, $n-1$ DBS would be needed. Consider the scheme shown in Fig.~\ref{fig3}(b) and the following four individual biphoton ququarts at the input modes of the DBS set:
%%%%%%%%%%%%%%%%
\begin{equation}
\begin{split}
|\psi\rangle_a & = |0\rangle_a,\\
|\psi\rangle_b & = |0\rangle_b + |3\rangle_b,\\
|\psi\rangle_c & = |0\rangle_c + |3\rangle_c,\\
|\psi\rangle_c & = |0\rangle_d.
\end{split}
\end{equation}
%%%%%%%%%%%%%%%%
The output state is found to be
%%%%%%%%%%%%%%%%%%
\begin{equation}
\begin{split}
|\psi\rangle_{out} & = |0\rangle_{a'} |0\rangle_{b'} |0\rangle_{c'} |0\rangle_{d'} 
+ |0\rangle_{a'} |1\rangle_{b'} |0\rangle_{c'} |2\rangle_{d'}  \\
&+ |2\rangle_{a'} |0\rangle_{b'} |1\rangle_{c'} |0\rangle_{d'} 
+ |2\rangle_{a'} |1\rangle_{b'} |1\rangle_{c'} |2\rangle_{d'},
\end{split}
\end{equation}
%%%%%%%%%%%%%%%%%
which is a rather simple looking four-ququart entangled state. 

%%%%%%%%%%%%%%%%%%
%%%%%%%%%%%%%%%%%%

To generate a more complex four-ququart entangled state, one simply needs to change one of the input states to a superposition state. For example, for the following biphoton ququarts entering the DBS set shown in Fig.~\ref{fig3}(b),
%%%%%%%%%%%%%%%%
\begin{equation}
\begin{split}
|\psi\rangle_a & = |0\rangle_a + |3\rangle_a,\\
|\psi\rangle_b & = |0\rangle_b,\\
|\psi\rangle_c & = |0\rangle_c + |3\rangle_c,\\
|\psi\rangle_c & = |0\rangle_d + |3\rangle_d,
\end{split}
\end{equation}
%%%%%%%%%%%%%%%%
the output state is calculated to be
%%%%%%%%%%%%%%%%%%
\begin{equation}
\begin{split}
|\psi\rangle_{out} & = |0\rangle_{a'} |0\rangle_{b'} |0\rangle_{c'} |0\rangle_{d'} 
+ |0\rangle_{a'} |0\rangle_{b'} |2\rangle_{c'} |1\rangle_{d'}  \\
&+ |1\rangle_{a'} |2\rangle_{b'} |0\rangle_{c'} |0\rangle_{d'} 
+ |1\rangle_{a'} |2\rangle_{b'} |2\rangle_{c'} |1\rangle_{d'} \\
&+ |2\rangle_{a'} |0\rangle_{b'} |1\rangle_{c'} |0\rangle_{d'} 
+ |2\rangle_{a'} |0\rangle_{b'} |3\rangle_{c'} |1\rangle_{d'} \\
&+ |3\rangle_{a'} |2\rangle_{b'} |1\rangle_{c'} |0\rangle_{d'} 
+ |3\rangle_{a'} |2\rangle_{b'} |3\rangle_{c'} |1\rangle_{d'}.
\end{split}
\end{equation}
%%%%%%%%%%%%%%%%%

What we have shown in this section is that, with a relatively simple set of biphoton ququarts, it is possible to generate interesting-looking postselected entangled states of multiple ququarts. Scaling up to multiple ququarts becomes more or less a practical concern.

%%%%%%%%%%%%%%%%%
\section{Discussion}
%%%%%%%%%%%%%%%%%%

We have investigated scalable linear optical schemes to generate entanglement of multiple ququarts, in which the individual ququarts are constructed with the biphoton polarization states of frequency-nondegenerate SPDC. We have shown that it is possible to generate various entangled states of multiple ququarts by interfering them at beam splitter, such as, BS, PBS, and DBS. 

It is now relavant and interesting to discuss the efficiency of beamsplitter entangling scheme. Especially, we would like to know which kind of beamsplitter is best suited for generating an entangled state of multiple ququarts. 

Consider, first, the output two-ququart state of a BS shown in eq.~(\ref{BS_final}). As discussed in section \ref{sec:bs}, when the input state is of the most general form, there are total of 448 additional biphoton amplitudes that are omitted due to the two-ququart detection postselection. Experimentally, this means that there are simply too many amplitudes that do not result in four-fold coincidence so the experiment will be extremely inefficient in terms of the observable four-fold coincidence rate versus the SPDC pump power. 

The situation is much better with the PBS output state shown in eq.~(\ref{PBSout4}) as there are only 36 biphoton amplitudes that cannot be expressed, hence not shown, in the ququart basis states defined in eq.~(\ref{eq1}). With the DBS, the double-pair terms in eq.~(\ref{real_in2}) do not contribute any detectable amplitudes to the postselected output state in eq.~(\ref{dbs}). These omitted terms add up to total of 32 amplitudes and they cannot be detected by the two-ququart detection scheme. 

From the experimental point of view, it is best if there are no wasted amplitudes, i.e., the amplitudes that are thrown out by the four-fold coincidence circuit. This goal clearly is unachievable with SPDC sources but we may choose to improve the entangled state generation efficiency as much as possible by choosing to work with either the PBS or the DBS entangling scheme as both schemes offer roughly the same postselected-to-wasted amplitude ratio ($1/3$ for the PBS versus $1/2$ for the DBS). One additional advantage of the DBS scheme would be that, with postselection, the input state could be considered as the ideal input as in eq.~(\ref{input}) since the double-pair terms do not affect the output state.

Finally, we note that, although the ideal input state of the form in eq.~(\ref{eq3}) is not achievable with SPDC, it may be possible to use recently reported semiconductor quantum-dot photon pair sources with supressed multi-pair events to prepare such an ideal input state \cite{akopian}. In this case, the non-postselection nature of the DBS entanglement scheme could prove to be important.

%%%%%%%%%%%%%%%%%
\acknowledgments
%%%%%%%%%%%%%%%%%

We wish to thank J.P. Dowling and S.P. Kulik for helpful comments. This work was supported, in part, by the Korea Research Foundation (R08-2004-000-10018-0 and KRF-2006-312-C00551), the Korea Science and Engineering Foundation (R01-2006-000-10354-0), and POSTECH. 

%%%%%%%%%%%%%%%%%%%%%%%%%%%%%%%%%%%%%%


\begin{thebibliography}{}

\bibitem{ou} Z.Y. Ou \textit{et al.}, \prl \textbf{68}, 3663 (1992).

\bibitem{furusawa} A. Furusawa \textit{et al.}, Science \textbf{282}, 5389 (1998).

\bibitem{ralph} T.C. Ralph, \pra \textbf{61}, 010303 (1999). 

\bibitem{kas} D. Kaslikowski \textit{et al.}, Phys. Rev. Lett. \textbf{85}, 4418 (2000).

\bibitem{Popescu} D. Collins \textit{et al.}, Phys. Rev. Lett. \textbf{88}, 040404 (2002).

\bibitem{Chen} L.-B. Fu, J.-L. Chen, and S.-G. Chen, Phys. Rev. A. \textbf{69}, 034305(2004).

\bibitem{peres} H. Bechmann-Pasquinucci and A. Peres, Phys. Rev. Lett. \textbf{85}, 3313 (2000).

\bibitem{lee} S.-W. Lee, J. Ryu, and J. Lee, J. Korean Phys. Soc. \textbf{48}, 1307 (2006).

\bibitem{shih} Y.-H. Shih and C.O. Alley, \prl \textbf{61}, 2921 (1988).

\bibitem{kim} Y.-H. Kim, \pra \textbf{67}, 040301(R) (2003). 

\bibitem{gisin} J. Brenden \textit{et al.}, \prl \textbf{82}, 2594 (1999).

\bibitem{chek} M.V. Chekhova \textit{et al.} \pra \textbf{70}, 053801 (2004). 

\bibitem{bog} Yu.I. Bogdanov \textit{et al.} \prl \textbf{93}, 230503 (2004).

\bibitem{kulik3} Yu.I. Bogdanov \textit{et al.}, Phys. Rev. A. \textbf{73}, 063810 (2006).

\bibitem{kulik4} E.V. Moreva \textit{et al.}, Phys. Rev. Lett. \textbf{97}, 023602 (2006).

\bibitem{zeilinger1} A. Mair \textit{et al.} Nature (London) \textbf{412}, 313 (2001).

\bibitem{zeilinger2} A. Vaziri, G. Weihs, and A. Zeilinger, Phys. Rev. Lett. \textbf{89}, 240401 (2002).

\bibitem{howell} M.N. O'Sullivan-Hale \textit{et al.}, \prl \textbf{94}, 220501 (2005). 

\bibitem{padua} L. Neves \textit{et al.}, \prl \textbf{94}, 100501 (2005).

\bibitem{gisin2} H. de Riedmatten \textit{et al.} \pra \textbf{69}, 050304(R) (2004); R. T. Thew \textit{et al.}, Phys. Rev. Lett. \textbf{93}, 010503 (2004).

\bibitem{baek} S.-Y. Baek and Y.-H. Kim, \pra 75, 034309.(2007).

\bibitem{klyshko} D.N. Klyshko, \textit{Photons and Nonlinear Optics} (Gordon \& Breach, Amsterdam, 1988).

\bibitem{kok} P. Kok and S.L. Braunstein, \pra  \textbf{61}, 042304 (2000).

\bibitem{kim2} Y.-H. Kim, S.P. Kulik, and Y. Shih, Phys. Rev. A. \textbf{63}, 060301(R) (2001).

\bibitem{sim} Of course, there are many candidates for single-photon sources. If an efficient and reliable single-photon source becomes available, there would be no reason to use weak coherent states for quantum cryptography. For a review on single-photon sources, see B. Lounisand and M. Orrit, Rep. Prog. Phys. \textbf{68}, 1129 (2005). 

\bibitem{su21} V.P. Karassiov and S.P. Kulik, JETP, \textbf{104}, 30 (2007).

\bibitem{su22} G. Bj\"{o}rk \textit{et al.}, Proc. SPIE \textbf{4750}, 1 (2002).

\bibitem{james} D.F.V. James \textit{et al.}, \pra \textbf{64}, 052312 (2001), 

\bibitem{hom} C.K. Hong, Z.Y. Ou, and L. Mandel, \prl \textbf{59}, 2044 (1987).

\bibitem{sa} Y.H. Shih and C.O. Alley, \prl \textbf{61}, 2921 (1988).

\bibitem{akopian}  N. Akopian \textit{et. al.}, \prl \textbf{96}, 130501 (2006). 

\end{thebibliography}
\end{document}